\documentclass[twocolumn]{aastex701}
\usepackage{amsmath,amssymb}
\usepackage{bm}

\usepackage{enumitem}
\begin{document}

\title{\texttt{Nii-MALA}: A Fast Metropolis-Adjusted Langevin Sampler with Radial Velocity Benchmarks}
\correspondingauthor{Sheng Jin}
\email{jins@ahnu.edu.cn}

\author{Jian Jiao}
\affiliation{Department of Physics, Anhui Normal University, Wuhu 241002, China}
\affiliation{Center for Astrophysics and Astronomical Technology, Anhui Normal University, Wuhu 241002, China}
\email{1765911894@qq.com}

\author[0000-0002-9063-5987]{Sheng Jin}
\affiliation{Department of Physics, Anhui Normal University, Wuhu 241002, China}
\affiliation{Center for Astrophysics and Astronomical Technology, Anhui Normal University, Wuhu 241002, China}
\email{jins@ahnu.edu.cn}

\author{Wenxin Jiang}
\affiliation{Department of Statistics and Data Science, Northwestern University, Evanston, IL 60208, USA}
\email{wjiang@northwestern.edu}

\author[0000-0001-9424-3721]{Dong-Hong Wu}
\affiliation{Department of Physics, Anhui Normal University, Wuhu 241002, China}
\affiliation{Center for Astrophysics and Astronomical Technology, Anhui Normal University, Wuhu 241002, China}
\email{wudonghong@ahnu.edu.cn}

\begin{abstract}
Markov chain Monte Carlo is widely used for model assessment and parameter fitting in astronomy and astrophysics, leading to numerous ready-to-use packages implementing various sampling techniques. For large-scale analyses involving many individual targets, computational efficiency becomes paramount, as it dictates the overall runtime. We introduce \texttt{Nii-MALA}, a high-performance C implementation of the Metropolis-Adjusted Langevin Algorithm, which leverages Message Passing Interface for parallelization and incorporates automatic differentiation backends for gradient evaluation. The code offers flexible control via a configuration file, allowing users to specify the proposal step sizes for all model parameters across parallel chains. We validate its effectiveness on 15- and 30-dimensional Gaussian distributions and a bimodal distribution, and further benchmark it against several other codes using radial velocity data from 51 Pegasi b. 
Our benchmarks confirm the efficiency of Langevin sampling over random-walk sampling in obtaining the effective sample size, particularly as the dimensionality of the target function increases. This algorithmic advantage can be effectively leveraged when paired with an efficient automatic differentiation library. However, for complex functions commonly encountered in astronomy---such as radial velocity orbital fitting---the derivative-dependent Langevin sampler shows lower efficiency than parallel tempering or alternative sampling approaches, as revealed by effective sample size analysis.

\end{abstract}

\keywords{
Algorithms --
Computational methods   --
Radial velocity
}
\section[Introduction]{Introduction}
\label{sec:intro}

Markov chain Monte Carlo (MCMC), with an appropriate choice of transition algorithm, can produce a stationary distribution that closely approximates the target distribution after a sufficient number of sampling steps \citep{Metropolis1953,Hastings1970}.
As a result, it is now a cornerstone of modern astronomical data analysis for both model fitting and parameter estimation.
Specifically, for a given astronomical problem, Bayesian inference is employed to derive the posterior distribution, which is then sampled via MCMC. The resulting samples are subsequently used for model selection or parameter inference \citep{Sharma2017}.
However, MCMC encounters substantial difficulties in high-dimensional settings. In such spaces, the high-density posterior regions typically occupy only a small fraction of the full parameter volume, which causes naive random-walk Metropolis (RWM) algorithms to exhibit extremely slow mixing \citep{Roberts2001,Mattingly2012}.
Moreover, real-world problems frequently feature multimodal posteriors, in which the basic RWM sampler can easily become stuck in a single mode \citep{L2025}.

Astronomy applications frequently encounter multi-parameter posterior distributions. For instance, orbital parameter retrieval for a single-planet system typically requires 7 or more parameters \citep{Gregory2005,PenaJenkins2025,Tan2026}, while a two-planet system roughly doubles this number to 13-15 parameters or more \citep{Huang2025,Jia2026,Jin2026}. To address these challenging posteriors, a variety of specialized sampling algorithms and software packages have been developed.
These include the ensemble affine sampler \texttt{emcee} and its derivatives \texttt{ptemcee} and \texttt{reddemcee} \citep{ForemanMackey2013,VousdenFarrMandel2016,Pena2026}; nested sampling methods, which directly provide Bayesian evidence \citep{Skilling2004,Mukherjee2006,Feroz2009,Speagle2020}; and adaptive or automatic parallel tempering MCMC (APT-MCMC) schemes \citep{Gregory2005,Jin2022,JinJiangWu2024,Pena2026}. Collectively, these powerful methods and tools have proven highly effective in solving challenging sampling tasks in modern astronomy.
For exoplanet transit-timing variation (TTV) models, Tuchow et al. \citep{TuchowFordPapamarkouLindo2019} compared the sampling efficiency of several MCMC methods, including the Metropolis-Adjusted Langevin Algorithm (MALA), Hamiltonian Monte Carlo (HMC), and affine-invariant ensemble samplers.
Although these methods and tools have proven highly successful for analyzing individual targets, the upcoming deluge of data from various missions---such as the high-precision astrometric time-series data from Gaia \citep{Perryman2014,Lammers2026}---necessitates a re-evaluation of these codes, particularly regarding their efficiency in batch analyses of large numbers of targets.

The overall sampling efficiency should be assessed from two complementary perspectives.
First, algorithmic sampling efficiency. For instance, optimal scaling theory shows that, for a broad class of high-dimensional targets, the RWM algorithm is most efficient when tuned to an average acceptance rate of approximately 0.234 \citep{Gelman1996,RobertsGelmanGilks1997}.
In contrast, derivative-based algorithms can achieve substantially higher acceptance rates. For example, MALA, which uses gradient information, has an optimal acceptance rate of about 0.574 for high-dimensional targets \citep{RobertsRosenthal1998,PillaiStuart2012}, while HMC reaches an optimal acceptance rate of 0.65 \citep{Neal2011}. The No-U-Turn Sampler (NUTS) adaptively sets the HMC trajectory length \citep{Hoffman2014}. These acceptance rates are tuning diagnostics rather than direct measures of sampling efficiency.
Second, computational implementation efficiency. Derivative-based methods such as MALA and HMC require extra computations to evaluate gradients, introducing a trade-off that depends on the effective sample size (ESS), autocorrelation, and the cost of the automatic differentiation backend \citep{Neal2011,Hoffman2014,Livingstone2022}.
For a retained chain of length $N$, the ESS estimates the number of independent draws represented by the correlated samples, while the integrated autocorrelation time (IAT) gives the corresponding correlation penalty, with $\mathrm{ESS}\approx N/\mathrm{IAT}$.
Meanwhile, practical implementations of MALA have continued to develop. \texttt{Pigeons.jl} supports parallel tempering with MALA \citep{SurjanovicEtAl2025}, while autoMALA adaptively adjusts the MALA step size \citep{BironLattesEtAl2024}.
Aiming to develop a highly efficient sampling code for  batch fitting of a large number of targets, we present \texttt{Nii-MALA}\footnote{\url{https://github.com/jianjiao-ah/Nii-MALA}}, a fast C-language MALA implementation.
It combines the message-passing interface (MPI) parallelization  framework of the \texttt{Nii-C} code \citep{JinJiangWu2024} with automatic differentiation through ADOL-C \citep{GriewankJuedesUtke1996,GriewankWalther2008} or the compiler-based Enzyme backend \citep{MosesChuravy2020}.
Furthermore, \texttt{Nii-MALA} offers flexible control via a configuration file, allowing users to manually specify proposal scales for all model parameters across parallel chains.
We first evaluated its performance on controlled 15- and 30-dimensional Gaussian targets and a bimodal distribution.
The MALA configurations achieve acceptance rates near 0.57, while RWM is near 0.23. However, acceptance rate alone does not fully reflect sampling efficiency. We therefore compare the methods using ESS, IAT, ${\rm ESS}/s$, which measures the effective sample size obtained per second of computation, and ${\rm ESS}/10^6$, which quantifies the effective sample size per one million sampling iterations at the cold temperature.

Furthermore, we carried out benchmarks of \texttt{Nii-MALA} with ADOL-C and Enzyme, \texttt{Nii-C} \citep{JinJiangWu2024}, \texttt{ptemcee} \citep{VousdenFarrMandel2016}, and \texttt{reddemcee} \citep{Pena2026} using the real-world radial velocity (RV) data of the exoplanet 51 Pegasi b \citep{MayorQueloz1995}.
The RV benchmark serves as a real-world comparison of the five tested samplers, using either MALA or RWM configurations. This comparison is intended only as a practical technical guideline for real-world tasks such as orbital parameter fitting in astronomy, with the goal of assisting practitioners in their choice of MCMC software and algorithms.

This paper is organized as follows. Section \ref{sec:method} details the implementation of MALA within the \texttt{Nii-MALA} code. Section \ref{sec:examples} presents benchmark results evaluating the code's effectiveness and the computational efficiency of the tested algorithms and codes.
Section \ref{sec:summary} concludes with a summary and discussion.

\section[MALA Implementation]{MALA Implementation}
\label{sec:method}

We consider a target density  $\pi$ on $\mathbb{R}^d$ and aim to draw samples from it. Within the Bayesian framework adopted here, $\pi$ represents the unnormalized posterior, which is proportional to the prior multiplied by the likelihood.

Derived from the Langevin stochastic differential equation, MALA is a gradient-based MCMC method whose dynamics feature both a deterministic drift (directed by the gradient of the log-target) and stochastic Brownian diffusion \citep{RosskyDollFriedman1978,RobertsTweedie1996}.
To correct for the bias induced by time-discretization, a Metropolis–Hastings accept/reject step is added, ensuring that the Markov chain preserves the target invariant distribution $\pi$ \citep{Metropolis1953,Hastings1970,RobertsRosenthal1998,PillaiStuart2012}.

Let $W_t$ be a standard $d$-dimensional Brownian motion, that is, a continuous stochastic process with independent Gaussian increments. The
Langevin It\^{o} diffusion associated with the target density $\pi$ is
defined by
\begin{equation}
  \mathrm{d}X_t
  = \nabla \log \pi(X_t)\, \mathrm{d}t
    + \sqrt{2}\, \mathrm{d}W_t,
  \label{eq:langevin-main}
\end{equation}
where $\nabla \log \pi$ denotes the gradient of the log-target density.
The It\^{o} interpretation specifies how the stochastic integral in Equation~\eqref{eq:langevin-main} is defined \citep{Oksendal2003}.
Under suitable regularity conditions, the diffusion in
\eqref{eq:langevin-main} admits $\pi$ as its invariant distribution
\citep{RobertsTweedie1996}.

Simulating \eqref{eq:langevin-main} requires discretizing the continuous-time dynamics. Applying the Euler--Maruyama scheme with step size $\tau>0$ yields, for $t=0,1,2,\dots$, the recursion
\begin{equation}
  X_{t+1}
  = X_t + \tau\, \nabla \log \pi(X_t)
        + \sqrt{2 \tau}\, \xi_t,
  \quad
  \xi_t \sim \mathcal{N}_d(0, I_d),
  \label{eq:ula-update}
\end{equation}
where $\xi_t$ are independent and identically distributed (i.i.d.) random vectors and $\mathcal{N}_d(0,I_d)$ denotes the $d$-dimensional standard Gaussian distribution with zero mean and identity covariance matrix $I_d$.

Given the current state $x = X_t$, a candidate state $x' = X_{t+1}$ is generated using Equation \eqref{eq:ula-update},
where $x'$ is distributed according to the proposal density  $q(x' \mid x)$, which is given explicitly (up to a normalizing constant)  by
\begin{equation}
  q(x' \mid x)
  \propto
  \exp\!\left(
    - \frac{1}{4 \tau}
      \left\|
        x' - x - \tau\, \nabla \log \pi(x)
      \right\|_2^2
  \right).
  \label{eq:ula-transition}
\end{equation}
Here, $\lVert\cdot\rVert_2$ denotes the Euclidean ($L_2$) norm.
In general, this unadjusted Langevin algorithm does not preserve $\pi$ exactly. This bias can be corrected by adding a  Metropolis--Hastings accept--reject step to the Langevin update
\citep{RobertsTweedie1996,RobertsRosenthal1998}, whereby the proposal $x'$ is
accepted with probability
\begin{equation}
  \alpha(x, x')
  = \min\left\{
      1,\;
      \frac{\pi(x')\, q(x \mid x')} {\pi(x)\, q(x' \mid x)}
    \right\}.
  \label{eq:mala-accept}
\end{equation}

The accept--reject decision is then made by drawing a random number
$u_t \sim \mathcal{U}(0,1)$ and comparing it with
$\alpha(x,x')$, where $\mathcal{U}(0,1)$ denotes the uniform distribution on $[0,1]$. If $\alpha(x, x')=1$, the proposal is  accepted deterministically; otherwise, it is accepted with probability $\alpha(x, x')$. Detailed balance means that the probability flow between any pair of states is equal in both directions. This condition makes the chain reversible and leaves $\pi$ unchanged by the transition, so that $\pi$ is the invariant distribution \citep{Tierney1994}.
For numerical stability, we evaluate the acceptance probability \eqref{eq:mala-accept} on the logarithmic scale to prevent numerical underflow for very small probabilities.

Moreover,  to provide users with greater flexibility and control, \texttt{Nii-MALA} runs multiple parallel chains via MPI-based parallelization, where each chain can either evolve independently or incorporate a parallel tempering function if desired.
Users can manually specify the proposal step sizes for individual parameters in each chain.
\texttt{Nii-MALA} determines the
proposal scale  of each parameter using three control parameters:  \texttt{init\_gp\_ratio} ($\rho$) , which sets the baseline initial proposal size relative to the prior range; \texttt{init\_step\_para} ($\eta$), a parameter-wise vector of length equal to the number of parameters;  and \texttt{scale\_step\_beta} ($\kappa$), a chain-wise vector of length equal to the number of parallel chains. For each parameter dimension $j$ in chain $r$, the
specific proposal scaling is defined as
\begin{equation}
  s_{r,j}
  = (\theta_j^{\max} - \theta_j^{\min}) \, \rho \, \eta_j \, \kappa_r ,
  \label{eq:init-scale-rj}
\end{equation}
where $s_{r,j}$ denotes the proposal scale for the $j$-th parameter in chain $r$,
$\eta_j$ is the $j$-th component of \texttt{init\_step\_para}, and $\kappa_r$  is the $r$-th component of \texttt{scale\_step\_beta},
$\theta_j^{\max}$ and $\theta_j^{\min}$ are the upper and lower bounds of the prior interval for parameter $j$, respectively.

The tested ADOL‑C implementation uses reverse mode (version 2.7.3). In its derivative calculations, it records a new tape at each likelihood–gradient call for the 2D, 15D, and 30D targets; for the RV benchmark, however, it records once and replays a single rank‑specific tape.
The tested Enzyme implementation uses Enzyme v0.0.289 with LLVM 19.1.1 \citep{MosesChuravy2020}. For the RV benchmark, both ADOL-C and Enzyme differentiate the likelihood and use a custom reverse rule for the iterative Kepler solver.

Adapting \texttt{Nii-MALA} to a new application requires modifying only two files: \texttt{user\_prior.c} for specifying the prior distributions and \texttt{user\_logll\_grad.cpp} for computing the log-likelihood function and its gradient with the selected automatic differentiation backend.

\section[Benchmarks]{Benchmarks}
\label{sec:examples}

This section presents a set of benchmarks to demonstrate the effectiveness of the \texttt{Nii-MALA} code.
Moreover, we compare the numerical performance of the tested codes and configurations.
Our goal is to provide a useful reference for future large-scale batch applications involving numerous targets, particularly where computational efficiency is paramount.
For the benchmarks, we ran parallel MCMC chains in each experiment and computed the ESS and IAT based on the cold chains. To evaluate performance, we recorded two metrics: ${\rm ESS}/s$ and ${\rm ESS}/10^6$.

The first subsection compares the standard RWM sampler with MALA. The second subsection compares the computational efficiency of five sampler configurations in an astronomical RV application.

\subsection{RWM versus MALA}

Although MALA uses gradient information to improve its proposals, calculating the gradient incurs additional computational costs. Here, we use 15- and 30-dimensional Gaussian distributions and a 2-dimensional bimodal distribution to compare the overall computational efficiency of RWM and MALA, and to explore whether this trade-off is worthwhile. Both samplers---RWM (\texttt{Nii-C}) and MALA (\texttt{Nii-MALA})---are implemented in C, with \texttt{Nii-MALA} using two different backends---ADOL-C and Enzyme---to test the computational efficiency of different automatic differentiation implementations.

For these synthetic benchmarks, the overall proposal scale was adjusted through \texttt{init\_gp\_ratio}. The fixed values were 0.002 (15D), 0.0019 (30D), and 0.0054 (2D) for MALA, and 0.0178 (15D), 1.05 (30D), and 0.06 (2D) for RWM. These values were set based on the acceptance rates of test runs.
% \texttt{init\_gp\_ratio}as explained above%

\subsubsection{15- and 30-dimensional Gaussian}

Our first example considers 15- and 30-dimensional Gaussian distributions of the form:
\begin{equation}
\pi(\mathbf{x}) \propto \prod_{i=1}^d \exp\left\{-(x_i - m_i)^2\right\}
\end{equation}
where $m_i$ is the center of the $i$-th coordinate, and we set the dimension to $d = 15$ and $30$.

\begin{figure*}[!htbp]
\centering
\begin{minipage}[t]{0.48\textwidth}
\centering
\includegraphics[width=\textwidth]{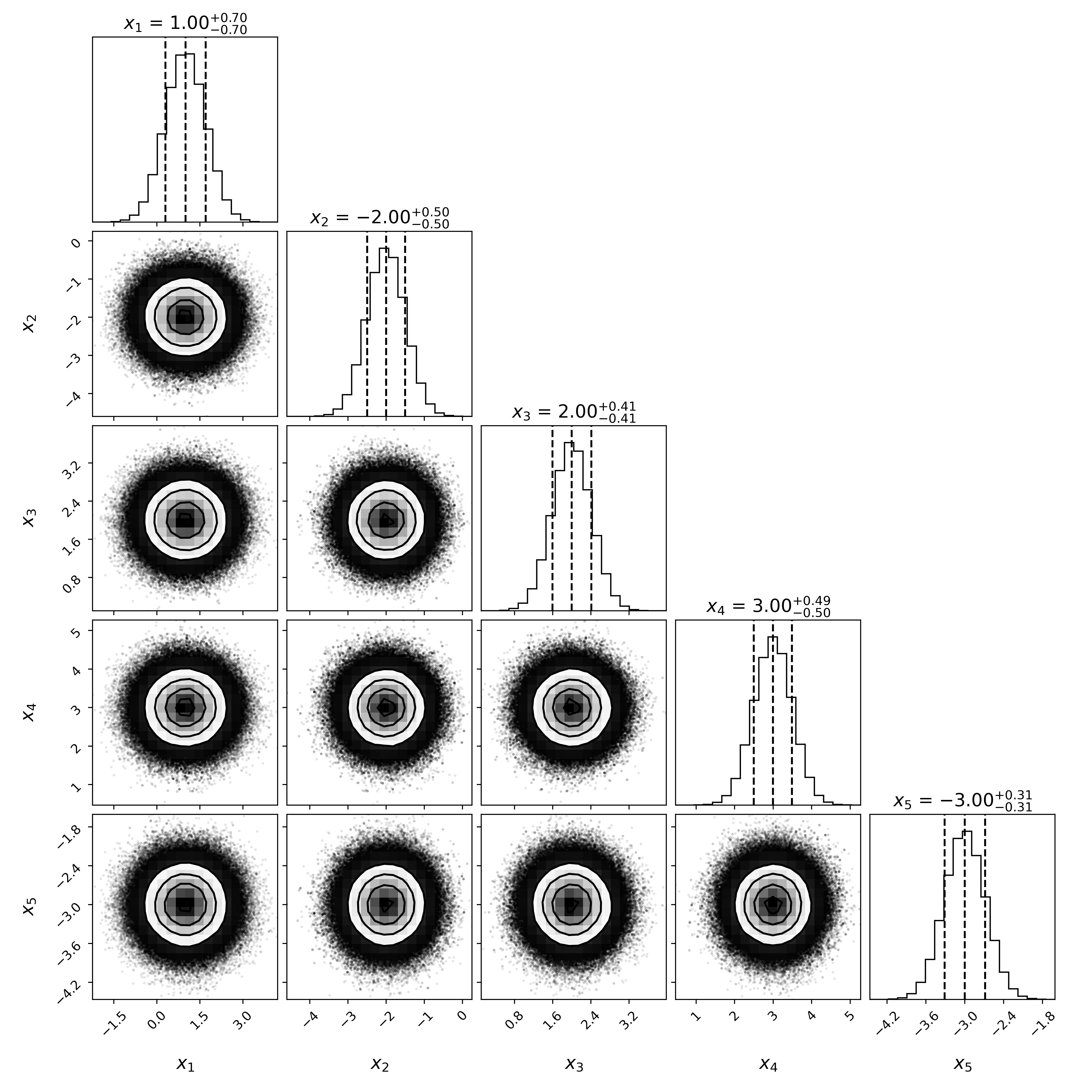}
\centerline{(a) \texttt{Nii-MALA}}
\end{minipage}
\hfill
\begin{minipage}[t]{0.48\textwidth}
\centering
\includegraphics[width=\textwidth]{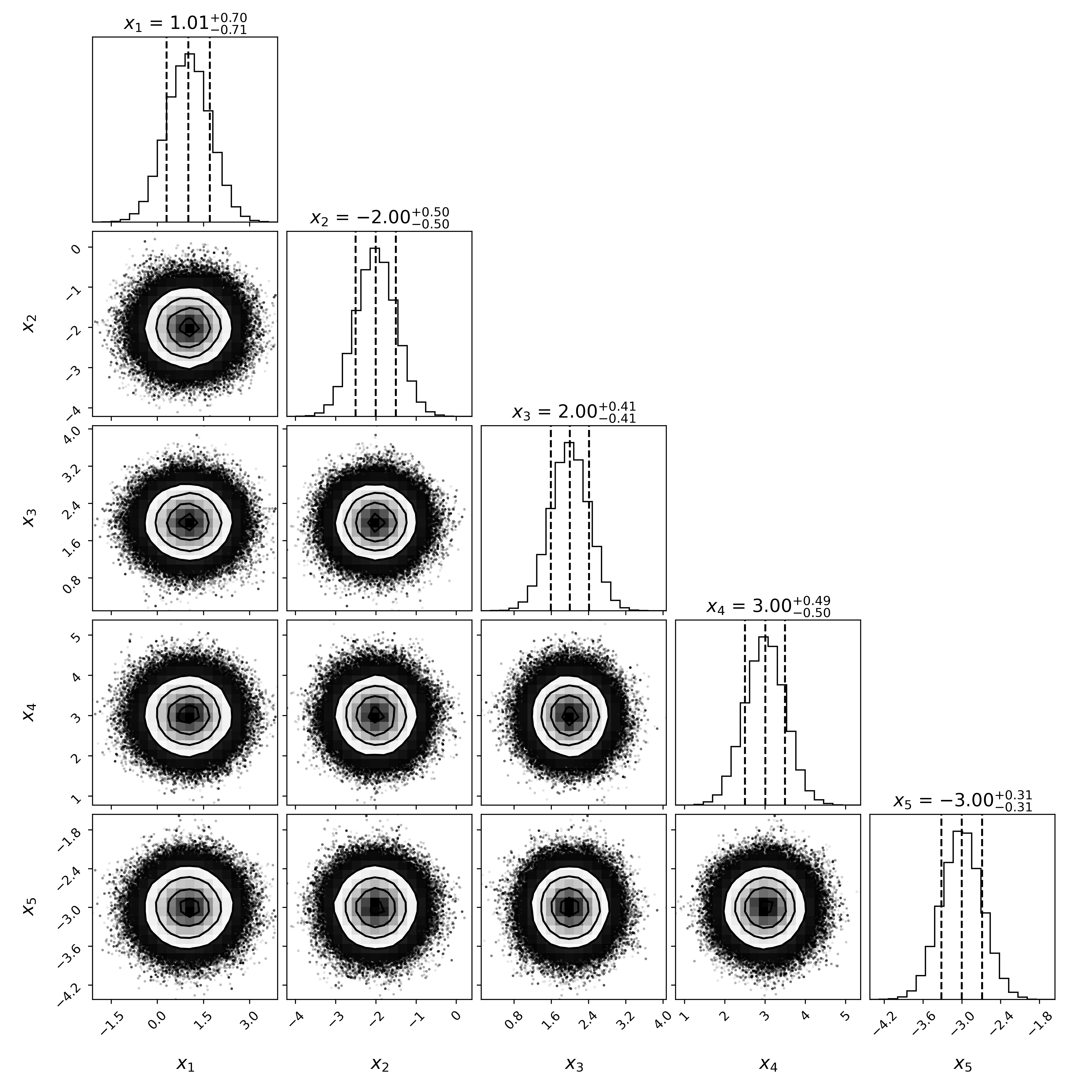}
\centerline{(b) \texttt{Nii-C}}
\end{minipage}
\caption{Corner plots for the representative $x_1$--$x_5$ subset of the 30-dimensional Gaussian benchmark, obtained with \texttt{Nii-MALA} (MALA) and \texttt{Nii-C} (RWM). Each chain used 1,000,000 iterations, and the first 200,000 iterations (20\%) of the cold chain were discarded as burn-in. Full 30-dimensional numerical diagnostics are reported in Table~\ref{tab:aptmcmc-mala-compare}.}
\label{fig:gaussian-30d-corner}
\end{figure*}

This simple target function serves as a controlled benchmark for comparing the sampling and computational efficiencies of the basic RWM and MALA settings. Each chain used 1,000,000 iterations, with the first 200,000 iterations of the cold chain discarded as burn-in.
For visual simplicity and to avoid an excessive number of contours, Figure~\ref{fig:gaussian-30d-corner} presents only the representative $x_1$--$x_5$  subset of the corner plots for \texttt{Nii-C} (RWM) and \texttt{Nii-MALA} (MALA), demonstrating that both codes recover the 30-dimensional Gaussian target.
Table~\ref{tab:aptmcmc-mala-compare} compares the sampling efficiency and runtime of the two codes.
All tests were carried out on a HASEE computer running Ubuntu 24.04.1 LTS with a 12th Gen Intel Core i7-12650H processor.
For both Gaussian dimensions, MALA achieves acceptance rates of approximately $57\%$, while RWM is near $22\%$.
For the Gaussian targets, MALA shows a clear ESS advantage over RWM, and this advantage becomes larger at the higher tested dimension. The median bulk ESS of MALA is about 2.8 times that of RWM in 15 dimensions and about 4.5 times that of RWM in 30 dimensions. The ADOL-C and Enzyme backends use the same MALA settings and give the same ESS values, but Enzyme requires much less runtime. As a result, Enzyme-assisted \texttt{Nii-MALA} achieves the highest median ${\rm ESS}/s$ for both Gaussian tests. These results show that the sampling advantage of MALA becomes more useful at the higher tested dimension when an efficient automatic differentiation backend is used.

\begin{table*}[!htbp]
\centering
\caption{Runtime and ESS-based efficiency for the synthetic benchmarks.}
\label{tab:aptmcmc-mala-compare}
\scriptsize
\resizebox{\textwidth}{!}{%
\begin{tabular}{llrrrrrrrr}
\hline
Model & Method & Iterations & Runtime (s) & Acceptance (\%) & Min. bulk ESS & Median bulk ESS & Max. IAT & Median ESS/s & ESS/$10^6$ \\
\hline
15D Gaussian & Nii-MALA (ADOL-C) & 1,000,000 & 11.662 & 56.951 & 18,743.28 & 42,080.57 & 42.68 & 3,608.35 & 42,080.57 \\
15D Gaussian & Nii-MALA (Enzyme) & 1,000,000 & 3.638 & 56.951 & 18,743.28 & 42,080.57 & 42.68 & 11,566.95 & 42,080.57 \\
15D Gaussian & Nii-C (RWM) & 1,000,000 & 3.526 & 22.535 & 5,151.17 & 14,938.05 & 155.30 & 4,236.54 & 14,938.05 \\
\hline
30D Gaussian & Nii-MALA (ADOL-C) & 1,000,000 & 20.699 & 57.079 & 17,698.51 & 39,553.81 & 45.20 & 1,910.90 & 39,553.81 \\
30D Gaussian & Nii-MALA (Enzyme) & 1,000,000 & 7.283 & 57.079 & 17,698.51 & 39,553.81 & 45.20 & 5,430.98 & 39,553.81 \\
30D Gaussian & Nii-C (RWM) & 1,000,000 & 6.935 & 22.129 & 8,159.57 & 8,697.05 & 98.04 & 1,254.08 & 8,697.05 \\
\hline
2D bimodal & Nii-MALA (ADOL-C) & 2,000,000 & 21.593 & 55.330 & 4,983.41 & 20,640.14 & 321.07 & 955.86 & 10,320.07 \\
2D bimodal & Nii-MALA (Enzyme) & 2,000,000 & 2.458 & 55.376 & 4,586.05 & 21,908.74 & 348.88 & 8,911.52 & 10,954.37 \\
2D bimodal & Nii-C (RWM) & 2,000,000 & 2.119 & 22.746 & 2,470.32 & 27,631.13 & 647.69 & 13,040.98 & 13,815.57 \\
\hline
\end{tabular}
}
\tablecomments{Iteration counts are reported per chain. ESS and IAT are calculated from the cold chain after discarding its first 20\% as burn-in. The final column reports median ${\rm ESS}/10^6$, normalized by the sampling iterations at the cold temperature.}
\end{table*}

\subsubsection{Bimodal distribution}

The second test case is a 2-dimensional bimodal distribution, comprising a Gaussian component and a curved, crescent-shaped component. Its target density is given by
\begin{align}
\pi(a,b) \propto{} & \frac{1}{2}\exp\left[-\frac{a^2}{1.44}-\frac{\left(b+1+\frac{4}{9}a^2\right)^2}{0.1225}\right] \nonumber \\
& + \frac{1}{2}\exp\left[-8\left(a^2+(b-2)^2\right)\right]
\end{align}

The first term forms the curved crescent through the quadratic dependence of its center in $b$ on $a$, whereas the second term is a compact Gaussian centered near $(0,2)$.

Parallel tempering was enabled in both \texttt{Nii-C} and \texttt{Nii-MALA} to facilitate movement between the two components \citep{JinJiangWu2024}.

\begin{figure*}[!htbp]
\centering
\begin{minipage}[t]{0.48\textwidth}
\centering
\includegraphics[width=\textwidth]{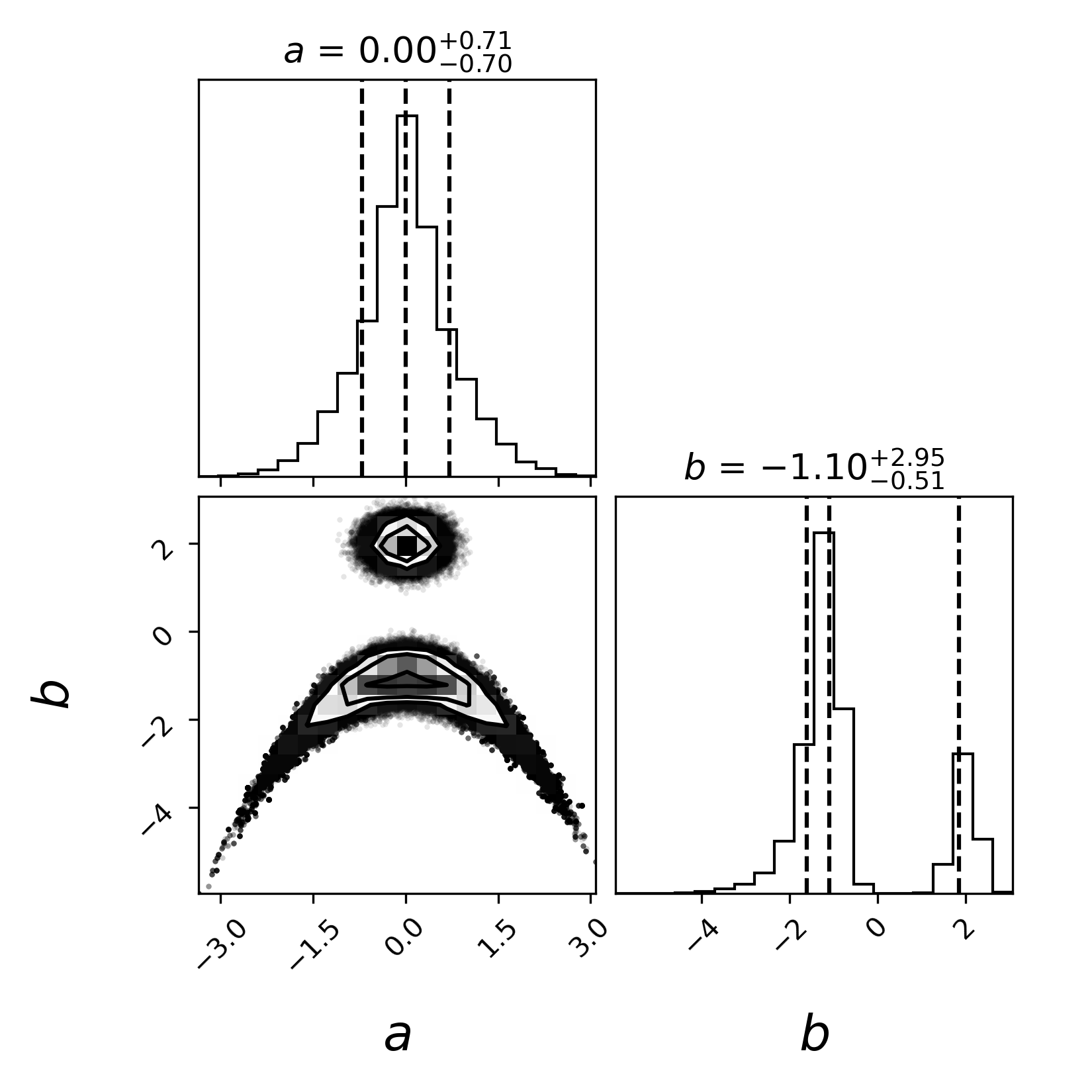}
\centerline{(a) \texttt{Nii-MALA}}
\end{minipage}
\hfill
\begin{minipage}[t]{0.48\textwidth}
\centering
\includegraphics[width=\textwidth]{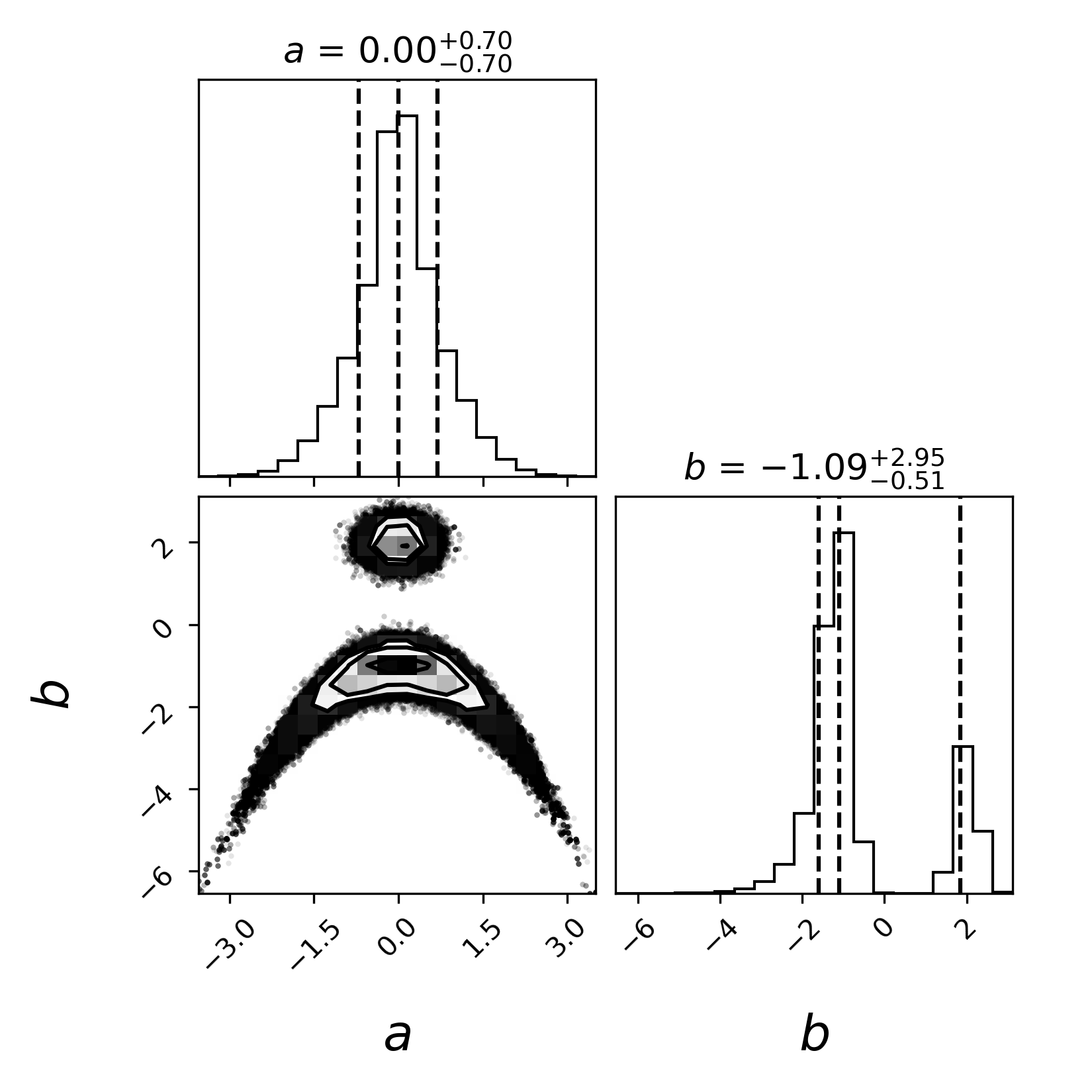}
\centerline{(b) \texttt{Nii-C}}
\end{minipage}
\caption{Corner plots for the two-dimensional bimodal benchmark, obtained using \texttt{Nii-MALA} (MALA) and \texttt{Nii-C} (RWM). Each chain used 2,000,000 iterations, and the first 400,000 iterations (20\%) of the cold chain were discarded as burn-in.}
\label{fig:bimodal-2d-corner}
\end{figure*}

Both \texttt{Nii-MALA} (MALA) and \texttt{Nii-C} (RWM) used 2,000,000 iterations per chain, with the first 400,000 iterations of the cold chain discarded as burn-in.
Figure~\ref{fig:bimodal-2d-corner} presents the corner plots of the 2-dimensional bimodal target obtained by \texttt{Nii-MALA} and \texttt{Nii-C}, demonstrating that both algorithms successfully recover the target bimodal distribution.

Table~\ref{tab:aptmcmc-mala-compare} compares the two samplers in terms of both algorithmic efficiency and computational runtime. Similar to the Gaussian benchmarks, MALA achieves an acceptance rate of $\sim$55\%, significantly higher than RWM's $\sim$23\%.
RWM has the highest median ${\rm ESS}/s$ for this target, followed by MALA (Enzyme) and MALA (ADOL-C). The median bulk ESS values are 20,640.14, 21,908.74, and 27,631.13 for MALA (ADOL-C), MALA (Enzyme), and RWM, respectively. These results show that RWM gives the highest median ESS and median ${\rm ESS}/s$ for this bimodal target.

The analytic mass of the crescent component is 0.770642. The estimated masses from MALA (ADOL-C), MALA (Enzyme), and RWM are 0.7870, 0.7707, and 0.7772, respectively.  The estimated component masses differ from the analytic value by approximately 1.64, 0.01, and 0.66 percentage points for MALA (ADOL-C), MALA (Enzyme), and RWM, respectively.

\subsection{Benchmarking RV Fitting}
\label{subsec:app-51peg-rv}

Here we benchmark the performance and runtime of five configurations using RV data for the exoplanet 51 Pegasi b \citep{MayorQueloz1995,Butler2006}, sourced from the NASA Exoplanet Archive \footnote{https://exoplanetarchive.ipac.caltech.edu}.

The purpose of this benchmark is to compare runtime and sampling efficiency in a real-world astronomy application. Since \texttt{Nii-MALA}, \texttt{Nii-C}, \texttt{ptemcee}, and \texttt{reddemcee} differ in their proposal mechanisms and software structures, this constitutes a practical runtime benchmark conducted within an actual application on a given machine---rather than an advanced programming-language study or a rigorous algorithmic analysis.
Accordingly, we set narrow prior ranges for fitting the RV of 51 Pegasi b, allowing all codes to rapidly locate the density mode and begin sampling in its vicinity.
This setup enables us to compare code efficiency in a relatively easy setting, where each code can readily recover the posteriors.

The implemented RV model comprises a total of 8 parameters:
\begin{equation}
\theta=(P,K,\lambda,h,k,\dot{\gamma},\gamma,\sigma_{\rm jit}),
\end{equation}
where \(P\) is the orbital period, \(K\) is the RV semi-amplitude,
\(\lambda\) is the mean longitude at the reference epoch, \(h=\sqrt{e}\sin\omega\),
\(k=\sqrt{e}\cos\omega\), \(\dot{\gamma}\) is a linear RV trend,
\(\gamma\) is the systemic velocity offset that accounts for the star's mean RV, and
\(\sigma_{\rm jit}\) is an additional white-noise jitter term.

The eccentricity and
argument of periastron are recovered as
\begin{equation}
e=h^2+k^2, \qquad \omega=\mathrm{atan2}(h,k),
\end{equation}
with the prior constrained to the physical range \(e<1\).

For each observation time \(t_i\), where
\(t_0\) is the first epoch in the data file, we evaluate the mean anomaly
\begin{equation}
M_i = \frac{2\pi(t_i-t_0)}{P}+\lambda-\omega,
\end{equation}
and then solve Kepler's equation \(E_i-e\sin E_i=M_i\) for the eccentric anomaly $E_i$, and compute the true anomaly
\(\nu_i\).

The corresponding RV signal at time $t_i$ is then given by
\begin{equation}
  v_{\rm mod}(t_i)=
  K\left[\cos(\nu_i+\omega)+e\cos\omega\right]
  +\dot{\gamma}(t_i-t_0)+\gamma .
  \label{eq:rv-model-51peg}
\end{equation}

Finally, the log-likelihood is given by
\begin{equation}
  \log L(\theta)=
  -\frac{1}{2}\sum_i
  \left[
    \log(2\pi s_i^2)+
    \frac{(v_i-v_{\rm mod}(t_i))^2}{s_i^2}
  \right].
  \label{eq:rv-loglike-51peg}
\end{equation}
where \(s_i^2=\sigma_i^2+\sigma_{\rm jit}^2\), with $\sigma_i$ denoting the measurement uncertainties of the observed RV $v_i$.

The priors of the 8 parameters are listed in Table \ref{tab:rv8-prior-settings}. We adopt uniform priors over relatively narrow ranges for
\(P,K,\lambda,h,k,\dot{\gamma}\), and \(\gamma\) to allow all MCMC codes to rapidly locate the posterior mode, despite their differing sampling algorithms.
The jitter prior is a truncated normal distribution with mean
\(5\,\mathrm{m\,s^{-1}}\) and standard deviation \(5\,\mathrm{m\,s^{-1}}\),
truncated to the interval \([10^{-5},20]~\mathrm{m\,s^{-1}}\).

\begin{table}[!htbp]
\centering
\caption{Prior distributions of the RV model.}
\label{tab:rv8-prior-settings}
\begin{tabular}{lccc}
\hline
Parameter & Prior & Min & Max \\
\hline
\(P\) (days) & Uniform & 3.0 & 5.0 \\
\(K\) (\(\mathrm{m\,s^{-1}}\)) & Uniform & 45.0 & 60.0 \\
\(\lambda\) (rad) & Uniform & 0 & \(2\pi\) \\
\(h=\sqrt{e}\sin\omega\) & Uniform & -1.0 & 1.0 \\
\(k=\sqrt{e}\cos\omega\) & Uniform & -1.0 & 1.0 \\
\(\dot{\gamma}\) (\(\mathrm{m\,s^{-1}\,day^{-1}}\)) & Uniform & -0.01369 & 0.00274 \\
\(\gamma\) (\(\mathrm{m\,s^{-1}}\)) & Uniform & -10.0 & 10.0 \\
\(\sigma_{\rm jit}\) (\(\mathrm{m\,s^{-1}}\)) & Truncated \(\mathcal{N}(5,5^2)\) & \(10^{-5}\) & 20.0 \\
\hline
\end{tabular}
\tablecomments{The parameters \(h\) and \(k\) are sampled
under the joint constraint \(h^2+k^2<1\), ensuring that the recovered
eccentricity satisfies \(e<1\).}
\end{table}

We use four codes, with a total of the five tested configurations: \texttt{Nii-MALA} with ADOL-C, \texttt{Nii-MALA} with Enzyme, \texttt{Nii-C}, \texttt{ptemcee}, and \texttt{reddemcee} \citep{JinJiangWu2024,VousdenFarrMandel2016,Pena2026}. \texttt{Nii-C} is the direct RWM baseline, while the two \texttt{Nii-MALA} configurations compare automatic differentiation backends within the same sampler. \texttt{ptemcee} provides an established ensemble parallel-tempering baseline, and \texttt{reddemcee} represents an astronomy-oriented tempering implementation.
All configurations used eight temperature levels and eight worker processes. For \texttt{Nii-C} and the two \texttt{Nii-MALA} runs, we employed eight MPI ranks and 2,000,000 iterations per chain, with the first 500,000 discarded as burn-in. For \texttt{ptemcee} and \texttt{reddemcee}, we used 50 walkers per temperature level, with each walker running 40,000 iterations and the first 10,000 discarded as burn-in.
These settings match the number of worker processes and the total iterations across the five configurations, providing a consistent basis for comparing their runtime and ESS-based sampling efficiency.

We also performed exploratory single-temperature tests with \texttt{emcee} \citep{ForemanMackey2013} and \texttt{BlackJAX} MALA \citep{Cabezas2024}. The \texttt{emcee} run used 50 walkers with 40,000 steps per walker and required approximately 342 seconds, while a single-chain \texttt{BlackJAX} MALA run required approximately 8 minutes.
Both tests failed to recover the correct posterior distribution, indicating that the standard settings of these two codes cannot solve the RV fitting problem. For this reason, we do not include these two tests in Table~\ref{tab:rv8-51peg-comparison}. It should be noted that these outcomes should be interpreted solely within the context of our RV fitting model, rather than as a conclusive assessment of overall code efficiency.

\begin{figure*}[!htbp]
  \centering
  \includegraphics[width=0.92\textwidth]{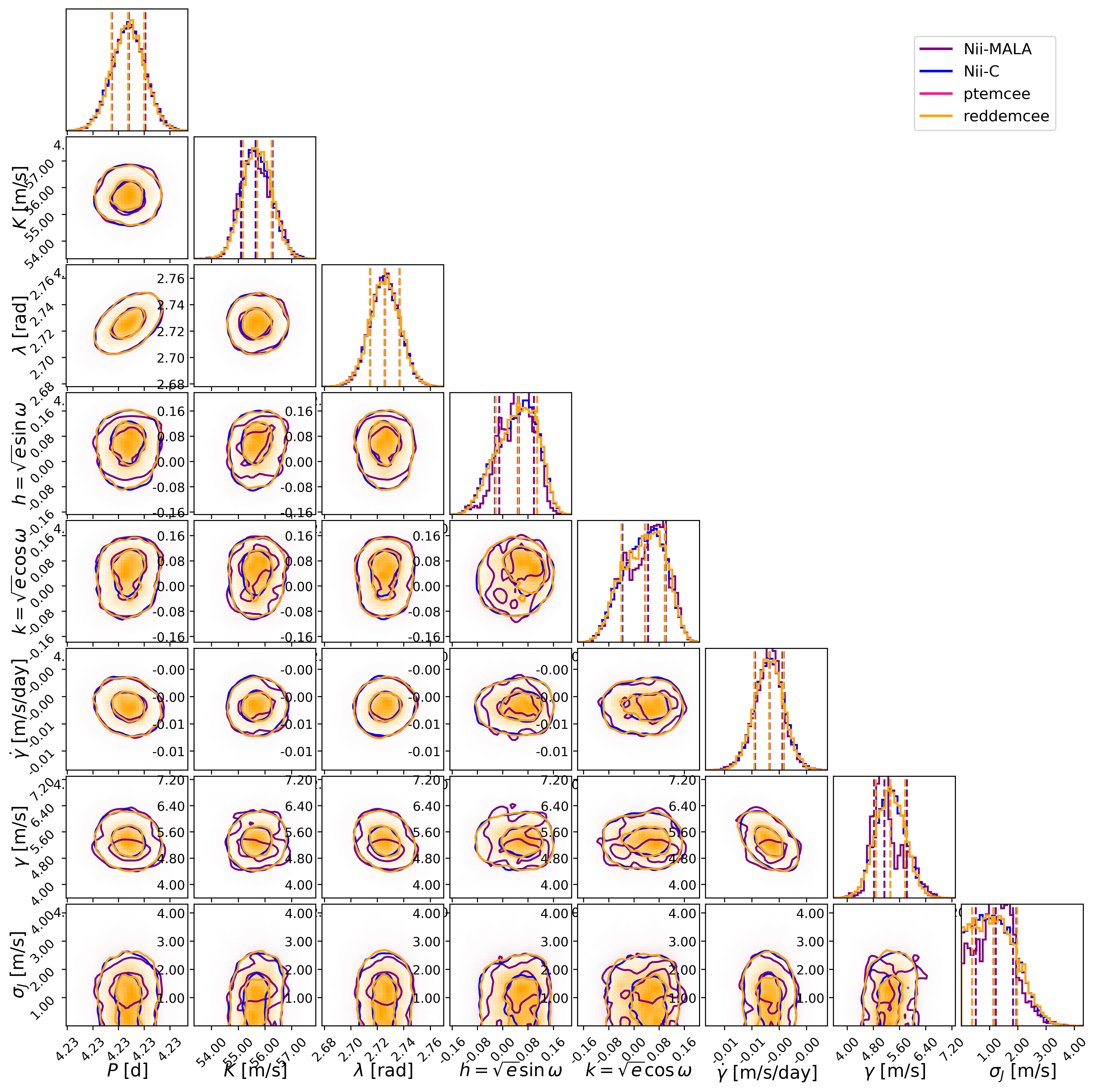}
  \caption{Superimposed corner plot of the eight RV-model parameters for 51 Pegasi b, inferred with \texttt{Nii-MALA} (ADOL-C and Enzyme), \texttt{Nii-C}, \texttt{ptemcee}, and \texttt{reddemcee}.}
  \label{fig:51peg-rv8-corner}
\end{figure*}

\begin{table*}[!htbp]
\centering
\caption{Comparison of the fitted major orbital parameters and sampling efficiency for 51 Pegasi b.}
\label{tab:rv8-51peg-comparison}
\scriptsize
\resizebox{\textwidth}{!}{%
\begin{tabular}{lccccrrr}
\hline
Sampler & $P$ (d) & $K$ (m s$^{-1}$) & $e$ & Runtime (s) &
Median ESS & Median ESS/s & ESS/$10^6$ \\
\hline

\texttt{Nii-MALA} (ADOL-C) &
$4.23078^{+0.00004}_{-0.00004}$ &
$55.694^{+0.544}_{-0.536}$ &
$0.0094^{+0.0094}_{-0.0065}$ &
3321.02 & 8,586.13 & 2.59 & 4,293.07 \\

\texttt{Nii-MALA} (Enzyme) &
$4.23078^{+0.00004}_{-0.00004}$ &
$55.709^{+0.548}_{-0.545}$ &
$0.0095^{+0.0095}_{-0.0066}$ &
406.01 & 7,715.81 & 19.00 & 3,857.91 \\

\texttt{Nii-C} &
$4.23078^{+0.00004}_{-0.00004}$ &
$55.677^{+0.567}_{-0.533}$ &
$0.0095^{+0.0095}_{-0.0066}$ &
216.05 & 46,480.28 & 215.14 & 23,240.14 \\

\texttt{ptemcee} &
$4.23078^{+0.00004}_{-0.00004}$ &
$55.707^{+0.544}_{-0.544}$ &
$0.0095^{+0.0096}_{-0.0066}$ &
408.47 & 46,745.18 & 114.44 & 23,372.59 \\

\texttt{reddemcee} &
$4.23078^{+0.00004}_{-0.00004}$ &
$55.709^{+0.548}_{-0.544}$ &
$0.0094^{+0.0095}_{-0.0066}$ &
696.93 & 35,897.21 & 51.51 & 17,948.61 \\
\hline
\end{tabular}
}
\tablecomments{Posterior constraints are the median and 16th--84th percentile interval. The first 25\% of the cold-temperature output was discarded before posterior and ESS calculations. Median ESS is calculated from $P$, $K$, and $e=h^2+k^2$; ${\rm ESS}/s$ divides this quantity by runtime. The final column reports median ${\rm ESS}/10^6$, normalized by the sampling iterations at the cold temperature.}
\end{table*}

All five configurations successfully recovered the orbital parameters of 51 Pegasi b, as shown in Figure~\ref{fig:51peg-rv8-corner}.
This figure shows superimposed corner plots obtained from MCMC sampling with the five configurations. The agreement among the posterior distributions confirms that all configurations recovered consistent parameters.
Table~\ref{tab:rv8-51peg-comparison} lists the retrieved orbital period, RV semi-amplitude, eccentricity, runtime, and ESS-based efficiency for each configuration. Only minor variations are observed in the posterior constraints.

In this benchmark, our main focus is the overall computational and sampling efficiency of the tested configurations in a real-world astronomy application.
Table~\ref{tab:rv8-51peg-comparison} also lists the recorded runtime.
All experiments were carried out on a HASEE computer running Ubuntu 24.04.1 LTS, equipped with a 12th Gen Intel Core i7-12650H processor and 16.0 GiB of memory.
The ADOL-C and Enzyme \texttt{Nii-MALA} runs require 3321.02 and 406.01 seconds, respectively.
The results show that the Enzyme implementation reduces the runtime of the ADOL-C implementation by a factor of 8.2, indicating that selecting a fast automatic differentiation library is critically important in MALA implementation.
The recorded runtimes for \texttt{Nii-C}, \texttt{ptemcee}, and \texttt{reddemcee} are 216.05, 408.47, and 696.93 seconds, respectively.
The corresponding median ${\rm ESS}/s$ values are 2.59, 19.00, 215.14, 114.44, and 51.51 for \texttt{Nii-MALA} (ADOL-C), \texttt{Nii-MALA} (Enzyme), \texttt{Nii-C}, \texttt{ptemcee}, and \texttt{reddemcee}. Thus, the lower Enzyme runtime improves MALA's code efficiency substantially relative to ADOL-C, but it does not change the ranking for this RV model.
The difference between the two MALA configurations shows that the measured runtime depends strongly on the specific automatic differentiation backend. 
The ${\rm ESS}/10^6$ metric leads to the same conclusion: 4,293.07 and 3,857.91 for the two MALA configurations, compared with 23,240.14, 23,372.59, and 17,948.61 for \texttt{Nii-C}, \texttt{ptemcee}, and \texttt{reddemcee}. 
Based on both ${\rm ESS}/s$ and ${\rm ESS}/10^6$, \texttt{Nii-C}, \texttt{ptemcee}, and \texttt{reddemcee} achieve higher ESS-based sampling efficiency than the two \texttt{Nii-MALA} configurations in this RV benchmark.
Thus, for the tested RV model, the use of gradient information does not provide an ESS-based efficiency advantage over the non-gradient samplers.

Note that the computational efficiency discussed here primarily refers to local posterior sampling and RV likelihood evaluation, as the prior ranges in our RV-fitting model are narrow and all codes are able to locate the posterior mode quickly.
A second aspect of practical sampling performance is global exploration---the ability of an MCMC algorithm to move efficiently between separated high-density regions in complex posterior spaces.
We do not explicitly address this aspect in the present benchmark, as it is highly sensitive to the specific problem and the structure of the posterior distribution, making it difficult to compare codes or draw general conclusions from a single test case.
%The RV result is confined to the stated model, priors, sampler configurations, differentiation backends, and numerical implementations.

\section[Summary]{Summary}
\label{sec:summary}

The first aim of this work is to present \texttt{Nii-MALA}, a MALA sampler implemented in C and leveraging an MPI parallel framework, designed to deliver a computationally efficient tool for Bayesian inference.
The code offers fine-grained control over the proposal scales of all parameters across chains, integrates support for parallel tempering, and supports for the ADOL-C and Enzyme automatic differentiation backends.
We validate the package through benchmarks on 15- and 30-dimensional Gaussian distributions and a bimodal distribution, alongside an RV-fitting problem, to demonstrate its effectiveness.

The second aim is to compare the computational efficiency of different sampling algorithms and implementations, thereby providing benchmark references for future large-scale target analyses.
Accordingly, we set up two benchmark suites to separately assess efficiency: one comparing MALA against RWM, and the other comparing five sampler configurations in an RV application. Our conclusions are as follows.

\begin{enumerate}[label=\Roman*.]
\item Derivative-dependent algorithms such as MALA achieve significantly higher acceptance rates than basic RWM sampling ($\sim$55--57\% vs. $\sim$22--23\%). For the 15- and 30-dimensional Gaussian targets, MALA also achieves higher ESS and lower autocorrelation than RWM, with its ESS advantage becoming more pronounced as the dimensionality increases. Enzyme-assisted \texttt{Nii-MALA} provides the highest median ${\rm ESS}/s$ in both Gaussian benchmarks, about three times that of the ADOL-C implementation. For the bimodal target, however, \texttt{Nii-C} achieves the highest median ${\rm ESS}/s$, while the MALA configurations show substantially more frequent transitions between the two components. Thus, the practical efficiency of MALA depends on both the target distribution and the computational cost of gradient evaluation.

\item Based on the RV-fitting test, all five sampler configurations in the formal comparison recover consistent posterior constraints for 51 Pegasi b. In terms of runtime, \texttt{Nii-C} is the fastest among the five tested configurations. Compared to \texttt{Nii-C}, \texttt{Nii-MALA} (Enzyme) and \texttt{ptemcee} run approximately 2 times slower, \texttt{reddemcee} runs about 3.5 times slower, and \texttt{Nii-MALA} (ADOL-C) runs about 15 times slower. In terms of sampling efficiency, \texttt{Nii-C}, \texttt{ptemcee}, and \texttt{reddemcee} achieve higher median ${\rm ESS}/s$ than the two \texttt{Nii-MALA} configurations, and the ${\rm ESS}/10^6$ values show the same pattern, with \texttt{Nii-C}, \texttt{ptemcee}, and \texttt{reddemcee} again outperforming the two \texttt{Nii-MALA} configurations. Thus, for the tested RV model, the use of gradient information does not provide an ESS-based efficiency advantage over the non-gradient samplers.

\item The RV benchmark also provides guidance on the practical use of derivative-dependent sampling algorithms such as MALA. The automatic differentiation backend has a major impact on computational performance: replacing ADOL-C with Enzyme reduces the \texttt{Nii-MALA} runtime from 3321.02 to 406.01 seconds, corresponding to an 8.2-fold speedup. Enzyme provides a similar improvement in the Gaussian benchmarks, reducing the MALA runtime to a level close to RWM while retaining the higher ESS of MALA. These results show that the choice of automatic differentiation library has a major impact on computational cost.

\end{enumerate}

In summary, \texttt{Nii-MALA}---with its C-language implementation---serves as a fast MALA code for real-world astronomical applications. 
The benchmarks show that gradient information can improve sampling efficiency, measured as ESS, but whether this advantage translates into shorter runtime depends on the specific characteristics of target distribution and the cost of gradient evaluation. When the improvement in mixing outweighs the gradient cost, MALA can provide higher computational efficiency, as demonstrated by the 15- and 30-dimensional Gaussian benchmarks; in other cases, non-gradient samplers can remain more efficient. 
Therefore, sampler selection for large-scale astronomical applications should be guided by the specific problem at hand, rather than by the mathematical sophistication or algorithmic complexity of the methods.

It should be noted that the computational efficiency discussed in this paper is confined to likelihood evaluation, as our RV benchmark employs narrow priors that enable all tested codes to readily locate the posterior modes.
As for mode-finding performance---which is a central objective of MCMC algorithms---this depends on the specific algorithm and the structure of the posterior, and is beyond the scope of this paper. In future work, we plan to investigate the mode-finding capabilities of various algorithms using more real-world examples, and to extend \texttt{Nii-MALA} with additional features to handle high-dimensional and challenging posterior distributions.

\begin{acknowledgments}
We thank the anonymous referee for their constructive comments, which have helped improve this paper. 
S.J. acknowledges support from the National Natural Science Foundation of China (Grant No. 11973094) and the Incubation Program of Anhui Normal University (2023GFXK153). D.W. acknowledges support from the National Natural Science Foundation of China (NSFC, Grant No. 12573076).
\end{acknowledgments}

\software{\texttt{Nii-MALA}, \texttt{Nii-C} \citep{JinJiangWu2024}, \texttt{ptemcee} \citep{VousdenFarrMandel2016},
\texttt{reddemcee} \citep{Pena2026}}

\bibliography{refs}
\bibliographystyle{aasjournalv7}

%% the collaboration and author truncation commands are used.


\end{document}